\documentclass[preprint,aps,showpacs]{revtex4}
\usepackage{epsfig}
\begin{document}

\title{Influence of tensor interactions on masses and decay widths of dibaryons}

\author{Hourong Pang}
\affiliation{Department of Physics, Nanjing University, Nanjing, 210093, P. R. China;\\
Institute of Theoretical Physics, Chinese Academy of Sciences,
Beijing, 100080, China}
\author{Jialun Ping}
\affiliation{Department of Physics, Nanjing Normal University,
Nanjing, 210097, P.R. China;\\
Center for Theoretical Physics, Nanjing University, Nanjing,
210093, P.R. China}
\author{Lingzhi Chen}
\affiliation{Department of Physics, Nanjing University, Nanjing,
210093, China}
\author{Fan Wang}
\affiliation{Center for Theoretical Physics and Department of Physics,
Nanjing University, Nanjing, 210093, P. R. China}
\author{T. Goldman}
\affiliation{Theoretical Division, Los Alamos National Laboratory,
Los Alamos, NM 87545, USA}

\begin{abstract}

The influence of gluon and Goldstone boson induced tensor interactions 
on the dibaryon masses and D-wave decay widths has been studied in
the quark delocalization, color screening model. The effective S-D
wave transition interactions induced by gluon and Goldstone boson
exchanges decrease rapidly with increasing strangeness of the channel. 
The tensor contribution of K and $\eta$ mesons is negligible in this 
model. There is no six-quark state in the light flavor world studied so far
that can become bound by means of these tensor interactions besides the 
deuteron. The partial D-wave decay widths of the $IJ^p=\frac{1}{2}2^+$ 
N$\Omega$ state to spin 0 and 1 $\Lambda\Xi$ final states are 12.0 keV 
and 21.9 keV respectively. This is a very narrow dibaryon resonance 
that might be detectable in relativistic heavy ion reactions by existing 
RHIC detectors through the reconstruction of the vertex mass of the
decay product $\Lambda\Xi$ and by the COMPAS detector at CERN 
or at JHF in Japan and the FAIR project in Germany in the future.

\end{abstract}


\maketitle

\section{Introduction}
There might be two kinds of dibaryon\cite{wang,prc51}. One is the
loosely bound type consisting of two octet baryons; the deuteron is a
typical example. The others are tightly bound; the $H$ particle
had been predicted to be such a six quark state although later
calculations cast doubt on it\cite{jaffe,pang}. Instead, 
a non-strange $IJ^p=03^+$ $d^*$ and a strangeness -6 $IJ^p=00^+$
di-$\Omega$ have been predicted to be tightly bound six quark
states, which are formed from decuplet
baryons\cite{prc39,prc51,mpla13, prc65,zhang,commun38}. The
strangeness -3 $IJ^p=\frac{1}{2}2^+$ N$\Omega$ has also been
predicted to be of the tightly bound type\cite{goldman,commun41,pang}.

The tensor interaction due to $\pi$ exchange plays a vital role in the
formation of loosely bound deuteron. In the $d^*$ case the 
tensor interaction contribution to its mass is minor but is critical 
for its D-wave decay to the NN final state\cite{ping}. There are other
near threshold and deeply bound dibaryon candidates found in two
systematic quark model calculations\cite{pang,zhang}. This naturally 
raises the question as to whether or not the tensor interaction adds 
sufficient strength to bind these other near threshold states to become 
strong interaction stable, as in the deuteron case? Conversely, is the 
tensor interaction weak enough to leave the high spin, deeply bound 
states as narrow dibaryon resonances, as was shown in the $d^*$ 
case\cite{ping}? 

The present calculation is aimed at answering these two questions for 
the dibaryon candidates in the u,d,s three flavor world within the 
extended quark delocalization, color screening model (QDCSM). Our 
results show that both the effective S-D wave transition interactions 
due to gluon and $\pi$ exchanges decrease rapidly with increasing 
strangeness, and that the tensor contributions of K and $\eta$ mesons 
are negligible after a short range truncation. Altogether, the tensor 
contributions are not strong enough to bind other near threshold
six quark states, such as the $SIJ^p=-401^+ \Xi\Xi$, to become strong
interaction stable with the sole exception of the deuteron. The D-wave 
decay widths of high spin, six quark states, such as the $SIJ^p=003^+ 
d^*$ and the $SIJ^p=-3\frac{1}{2}2^+ N\Omega$, are in the range of tens 
of MeV to tens of keV and so these states might be narrow dibaryon
resonances.

The extended QDCSM is briefly introduced in Section II. In
Section III, we present our results. The discussion and conclusion
are given in Section IV.

\section{Brief description of the extended QDCSM}

The QDCSM was put forward in the early 90's. Details can be found 
in Refs.\cite{prl69,prc51,ping}. Although the short range repulsion
and the intermediate range attraction of the $NN$ interaction are
reproduced by the combination of quark delocalization and
color screening, the effect of the long-range pion tail
is missing in the QDCSM. Recently, the extended QDCSM
was developed\cite{prc65}, which incorporates this long-range
tail by adding $\pi$-exchange but with a short-range cutoff
to avoid double counting because the short and intermediate range
interactions have been accounted for by the quark delocalization
and color screening mechanism\cite{phen}. The exchange of K and 
$\eta$ mesons has been shown to be negligible in this model 
approach\cite{commun38,pang}. Nevertheless, their effect, especially 
the tensor part, has been included in this calculation to check 
further whether they are negligible in our model approach.
The extended QDCSM not only reproduces the properties of the
deuteron well, but also improves agreement with NN scattering
data as compared to previous work\cite{lucpl03}.

The Hamiltonian of the extended QDCSM, wave functions and the
necessary equations used in the current calculation are given
below. The tensor interactions due to effective one gluon and 
the octet Goldstone boson exchanges are included.
The details of the resonating-group method (RGM) have
been presented in Refs.\cite{ping,Buchmann}.

The Hamiltonian for the 3-quark system is the same as the well
known quark potential model, the Isgur model. For the six-quark
system, we assume
\begin{eqnarray}
H_6 & = & \sum_{i=1}^6 (m_i+\frac{p_i^2}{2m_i})-T_{CM}
+\sum_{i<j=1}^{6}
    \left[ V_{conf}(r_{ij}) + V_G(r_{ij}) +V_{\pi}(r_{ij})
    \right] ,                    \nonumber \\
V_G(r_{ij}) & = & \alpha_s \frac{\vec{\lambda}_i \cdot
\vec{\lambda}_j }{4}
 \left[ \frac{1}{r_{ij}}-\frac{\pi}{2} \delta (\vec{r_{ij}})
 \left( \frac{1}{m^2_i}+\frac{1}{m^2_j}+\frac{4\vec{\sigma}_i
 \cdot \vec{\sigma}_j}{3m_im_j} \right)+\frac{1}{4m_im_jr_{ij}^3}
 S_{ij} \right], \nonumber  \\
V_{\pi}(r_{ij}) & = & \theta (r-r_0)
\frac{g_8^2}{4\pi}\frac{m_{\pi}^2} {4m_q^2} \frac{1}{r_{ij}}
e^{-m_{\pi} {r_{ij}}}\left[\frac{1}{3} \vec{\sigma}_i \cdot \vec{\sigma}_j
+Z(r_{ij})S_{ij}\right]\vec{\tau}_i\cdot \vec{\tau}_j,
   \label{hamiltonian} \\
S_{ij} & = &  3\frac{\vec{\sigma}_i \cdot \vec{r}_{ij} \vec{\sigma}_j
\cdot \vec{r}_{ij}}{r_{ij}^2}-\vec{\sigma}_i \cdot \vec{\sigma}_j,
\nonumber \\
Z(r) & = &  \frac{1}{3}+\frac{1}{m_{\pi}r}+\frac{1}{(m_{\pi}r)^2},
\nonumber \\
V_{conf}(r_{ij}) & = & -a_c \vec{\lambda}_i \cdot \vec{\lambda}_j
\left\{ \begin{array}{ll}
 r_{ij}^2 &
 \qquad \mbox{if }i,j\mbox{ occur in the same baryon orbit}, \\
 \frac{1 - e^{-\mu r_{ij}^2} }{\mu} & \qquad
 \mbox{if }i,j\mbox{ occur in different baryon orbits},
 \end{array} \right. \nonumber \\
\theta ({r_{ij}}-r_0) & = & \left\{
 \begin{array}{ll}  0 & \qquad r_{ij} < r_0, \\  1 & \qquad \mbox{otherwise},
 \end{array} \right. \nonumber
\end{eqnarray}
where $r_0$ is the short range cutoff for pion exchange between
quarks. $g_8$ is the $\pi$ quark coupling constant. $m_{\pi}$
is the measured $\pi$ mass. The K and $\eta$ meson exchange
interactions, which have not been shown explicitly in the above equation
but have been included in this calculation, have a form very similar to 
that for the $\pi$. The color screening constant, $\mu$, is to be determined 
by fitting the deuteron mass in this model. All other symbols have their 
usual meanings, and the confinement potential $V_{conf}(r_{ij})$ has been 
discussed in Refs.\cite{prc65,ping}.

The pion exchange interaction, $V_{\pi}(r_{ij})$, affects only the $u$
and $d$ quarks. We take these to have a common mass, $m_q=
m_d = m_u$, i.e., ignoring isospin breaking effects.

The quark wave function in a given nucleon (orbit) relative
to a reference center (defined by $\vec{S}$) is taken to
have a Gaussian form characterized by a size parameter, $b$,
\begin{equation}
\phi(\vec{r}-\vec{S})  =  \left( \frac{1}{\pi b^2} \right)^{3/4}
    e^{-\frac{1}{2b^2} (\vec{r} - \vec{S})^2} .  \label{qkwvfcn}
\end{equation}

The light quark mass, $m_q$, is chosen to be $\frac{1}{3}$ of the nucleon 
mass. The strange quark mass, $m_s$, baryon size parameter, $b$, effective
quark-gluon coupling constant, $\alpha_s$, and the strength of confinement, 
$a_c$, are all determined by reproducing the nucleon mass, the $\Delta-N$ 
mass difference, an overall fit to other ground state baryon masses and by 
requiring the nucleon mass to be variational stable with respect to its 
size parameter, b. The quark-pion coupling constant $g_8=g_{qq\pi}$ is 
obtained from the nucleon-pion coupling constant by a slight ($<10$\%) 
correction to the classic symmetry relation, viz.,
\begin{equation}
\frac{g^2_{NN\pi}}{4\pi} = (M_N/m_q)^2
    \left(\frac{5}{3}\right)^2 \frac{g^2_{8}}{4\pi} e^{m_{\pi}^2 b^2/2},
\end{equation}
where $M_N$ is the nucleon mass and the last factor provides
the correction due to the extent of the quark wavefunction in
the nucleon. The K and $\eta$ are assumed to have a flavor SU(3) symmetric
quark-meson coupling constant and the same short range cutoff, $r_0$,
as the $\pi$'s. The color screening parameter, $\mu$, has been
determined by matching our calculation to the mass of the
deuteron. All of the model parameters are listed in Table I.

\vspace{12pt} Table I: Model Parameters

\begin{tabular}{ccccccc} \hline\hline
$m_q, m_s(MeV)$&$b(fm)$&$a_c(MeV\cdot
fm^{-2})$&$\alpha_s$&$\frac{g^2_{8}}{4\pi}$&~~~$~~r_0(fm)$~~~~~~&$\mu(fm^{-2})$ \\
\hline
$313, 634$&~~~$0.6022$~~~&~~~$25.03$~~~& ~~~$1.5547
$~~~&$0.5926$&~~~$0.8$~~~&~~~$0.90$~~~
\\ \hline\hline
\end{tabular}

\vspace{12pt}

The model masses of all octet and decuplet baryons are listed
in Table II.

\vspace{12pt} Table II: Single Baryon Masses in Units of MeV

\begin{tabular}{ccccccccc} \hline\hline
&N&$\Sigma$&$\Lambda$&$\Xi$&$\Delta$&$\Sigma ^*$&$\Xi^*$&$\Omega$ \\ \hline
theor.&~~~ 939.0~~~&1217.5~~&~~1116.9~~&1357.6&~~1232.0~~
&~~1359.6~~&~~1499.7~~&~~1652.3~~ \\
expt.&~~~ 939~~~&1193~~&~~1116~~&~~1318~~&~~1232~~
&~~1385~~&~~1533~~&~~1672~~ \\ \hline\hline
\end{tabular}

\vspace{12pt} \noindent

We use the RGM to carry out a dynamical calculation.
The trial RGM di-baryon wave function is
\begin{equation}
\Psi(6q)={\cal A}\left [\psi_{B_1}(\xi_1)\psi_{B_2}(\xi_2) \right]^{IS}
\chi(\vec{R}),
\end{equation}
where $\cal A$ is the antisymmetrization operator,
$\psi_{B_i}(\xi_i)$ $i=1,2$ is the baryon internal wave
function including color-flavor-spin-orbital part, $[\cdots]^{IS}$
means coupling the individual color-isospin-spin into the channel
isospin-spin and overall color singlet.

To simplify the RGM
calculation, one usually introduces Gaussian functions with
different reference centers $S_i$ i=1...n, which play the role of
generating coordinates in this formalism, to expand the relative
motion wave function $\chi(\vec{R})$ of the two quark clusters,
\begin{equation}
\chi{(\vec R)} = (\frac{3}{2\pi b^2})^{3/4} \sum_i C_i
e^{-\frac{3}{4}(\vec R - \vec S_i )^2/b^2}.
\end{equation}

In principle, any set of base wave functions can be used to
expand the relative motion wave function. The choice of a
Gaussian with the same size parameter, $b$, as the single
quark wave function given in Eq.(\ref{qkwvfcn}), however,
allows us to rewrite the resonating group wave function as
a product of single quark wave functions; (see Eq.(\ref{multi})
below). This cluster wave function (physical basis) can be
expressed in terms of the symmetry basis, classified by the
symmetry properties in a group chain, which in turn allows
the use of group theory method to simplify the calculation
of the matrix elements of the six quark Hamiltonian\cite{wpg}.
In our calculations, we typically use 15 Gaussian functions to
expand the relative motion wave function over the range
0-9~fm. For near threshold channels, such as the deuteron and
H particle, more Gaussian functions are needed to extend the
boundary to a larger extent to obtain more precise results as 
we have done previously. But in this calculation we did not 
make that effort because it is not necessary for our purpose.

After including the wave function for the six-quark
center-of-mass motion, the ansatz for the two-cluster wave
function used in the RGM can be written as
%%%%%%%%%%%%%%%%%%%%%%%%%%%%%%%%%%%%%%%%%%%%%%%%%%%%%%%%%
\begin{eqnarray}
\Psi_{6q} & = & {\cal A} \sum_{i=1}^{n} \sum_k \sum_{L_k=0,2}
C_{i,k,L_k}
  \int d\Omega_{S_i}
  \prod_{\alpha=1}^{3} \psi_{R} (\vec{r}_{\alpha},\vec{S_i},\epsilon)
  \prod_{\beta=4}^{6} \psi_{L} (\vec{r}_{\beta},\vec{S_i} , \epsilon)  \nonumber \\
  & & [ [\eta_{I_{1k}S_{1k}}(B_{1k})\eta_{I_{2k}S_{2k}}(B_{2k})]^{IS_k}
  Y_{L_k}(\vec{S}_i)]^J [\chi_c(B_1)\chi_c(B_2)]^{[\sigma]}
    \label{multi}  ,
\end{eqnarray}
where $k$ is the channel index. For example, for $SIJ=-2, 0, 0$,
we have $k=1, 2, 3$, corresponding to the channels $\Lambda\Lambda$, 
N$\Xi$ and $\Sigma\Sigma$. An angular momentum projection has been 
applied for the relative motion and $L_k$ is the orbital angular 
momentum of the relative motion wave function of channel k.

The delocalized orbital wavefunctions, $\psi_{R}(\vec{r},\vec{S_i}
,\epsilon)$ and $\psi_{L}(\vec{r},\vec{S_i} ,\epsilon)$, are given by
\begin{eqnarray}
\psi_{R}(\vec{r},\vec{S_i}, \epsilon) & = & \frac{1}{N(\epsilon)}
    \left( \phi(\vec{r}-\frac{\vec{S_i}}{2}) + \epsilon
    \phi(\vec{r}+\frac{\vec{S_i}}{2}) \right) , \nonumber \\
\psi_{L}(\vec{r},\vec{S_i}, \epsilon) & = & \frac{1}{N(\epsilon)}
    \left(\phi(\vec{r}+\frac{\vec{S_i}}{2}) + \epsilon
    \phi(\vec{r}-\frac{\vec{S_i}}{2})\right) , \label{1q} \\
N(\epsilon) & = & \sqrt{1+\epsilon^2+2\epsilon e^{-S_i^2/4b^2}},
\nonumber
\end{eqnarray}
where $\phi(\vec{r}-\frac{\vec{S_i}}{2})$ and 
$\phi(\vec{r}+\frac{\vec{S_i}}{2})$ are the single-particle 
Gaussian quark wave functions referred to above in Eq.(\ref{qkwvfcn}), 
with different reference centers $\frac{{S_i}}{2}$ and $-\frac{{S_i}}{2}$, 
respectively. The delocalization parameter, $\epsilon$, is a variational 
parameter determined by the dynamics of the quark system
rather than being treated as an adjustable parameter.
%%%%%%%%%%%%%%%%%%%%%%%%%%%%%%%%%%%%%%%%%%%%%%%%%%%%%%%%%
The initial RGM equation is
\begin{equation}
\int H(\vec R, \vec{R'}) \chi (\vec{R'}) d\vec{R'} = E \int
N(\vec R,\vec{R'}) \chi (\vec{R'}) d\vec{R'} .
\label{RGM}
\end{equation}
With the above ansatz, the RGM Eq.(\ref{RGM}) is converted
into an algebraic eigenvalue equation,
\begin{equation}
\sum_{j,k,L_k} C_{j,k,L_k} H_{i,j}^{k',L_k',k,L_k}
  = E \sum_{j,k,L_k} C_{j,k,L_k} N_{i,j}^{k',L_k',k,L_k}
    \delta_{L_k',L_k},
   \label{GCM}
\end{equation}
where $N_{i,j}^{k',L_k',k.L_k}, H_{i,j}^{k',L_k',k,L_k}$ are 
the wave function overlaps and Hamiltonian matrix elements,
respectively, obtained for the wave functions of Eq.(\ref{multi}).

The partial width of a high spin dibaryon state decaying into a
D-wave BB final state is calculated using Fermi's golden rule, 
in its nonrelativistic approximation, of course. Final state
interactions have also been taken into account in our model
approach\cite{ping}. The decay width formula used in $N\Omega 
\rightarrow \Lambda\Xi$ is a little different from that given 
in \cite{ping} due to the fact that the decay products, $\Lambda$ 
and $\Xi$, are different particles with different masses,
\begin{eqnarray}
 \Gamma (N\Omega \rightarrow \Lambda\Xi)
& =& \frac{1}{2J+1} \sum_{M_{J_f},M_{J_i}} \frac{1}{4\pi^2} p \frac{\sqrt{(m^2_\Lambda+p^2)
(m^2_\Xi+p^2)}}{\sqrt{m^2_\Lambda+p^2}+\sqrt{m^2_\Xi+p^2}}
\int |M_{fi}|^2 d \Omega , \\
&& p=
\frac{\sqrt{(m^2_\Lambda-m^2_\Xi)^2+m^2_{N\Omega}
(m^2_{N\Omega}-2m^2_\Lambda-2m^2_\Xi)}}{2m_{N\Omega}}  .\nonumber
\end{eqnarray}

\section{Results}

Previously, we chose the di-$\Omega$ as an example to study
whether or not our model results were sensitive to the
meson-exchange cut-off parameter, $r_0$, and the result
demonstrated that they are not\cite{commun38}. Hence, we consider
it is sufficient to calculate six-quark systems of different
quantum numbers with a representative cutoff value
of $r_0=0.8 fm$. Table III displays the masses (in MeV) calculated
for the dibaryon states of interest here. The deuteron channel result
calculated previously is included in this table for comparison.
It should be noted that in our calculation we assume the
wavefunction to be zero at the boundary point,
which is the usual boundary condition for bound states; i.e.,
we always solve the RGM Eq.(\ref{GCM}) as an eigenvalue problem.
If the state is unbound, we will not obtain a stable minimum
eigenenergy in the course of extending the boundary point.
Therefore the "mass" listed in Table III for these unbound states
is not a true mass of a dibaryon state. However the
contributions of tensor interaction are still a meaningful measure.
The masses listed under the $Mass_{nt}$ and $Mass_{wt}$ are the
calculated masses of those channels without and with tensor interaction.

\begin{center}
Table III : Tensor Interaction Effect on the Masses of Six-Quark Systems
\begin{tabular}{cccc}  \hline\hline
S,I,J & channel& Mass$_{nt}$&Mass$_{wt}$ \\
\hline $0,0,1$ & $NN$ & $1880.295$ & $1876.770(1880.250)$ \\
$-2,1,1$ & $\Sigma\Sigma$ & $2418.686$ &  $2418.517 $ \\
$-3,1/2,1$ & $\Lambda\Xi$ & $2477.491 $&$2477.488$ \\
$-4,0,1$ &  $\Xi\Xi$ & $2718.190$ & $2718.189$    \\
$-5,1/2,1$ & $\Xi\Omega$ & $3012.167$ & $3012.140$ \\
\hline\hline
\end{tabular}
\end{center}

The first line of Table III is the deuteron channel. If the tensor
interaction is neglected the deuteron is unbound. Even if the
tensor effect of gluon exchange is included, the deuteron is still
unbound. (The calculated mass is shown in parentheses). The
tensor interaction due to $\pi$ exchange is critical to form the
actual stable deuteron and the $\eta$ meson contribution
is negligible. Fig. 1 shows the effective transition interactions
of the NN S-D coupling due to gluon and $\pi$ exchanges. The
$\eta$ contribution has not been shown because it is negligible.
Obviously, the $\pi$ contribution is dominant. The boundary point
is limited to 9 fm and coupling to $\Delta\Delta$ channels has not
been taken into account in this calculation; hence, the deuteron
mass is a little higher than the best fit one which we reported
previously. This weakness will not affect our conclusion regarding 
the tensor interaction effect mentioned above.

The $\Sigma\Sigma$ channel mass listed in the second line is lower
than its own theoretical threshold 2435 MeV but higher than the
N$\Xi$, $\Lambda\Sigma$ thresholds. A three S-wave channel
coupling calculation has been done. The lowest mass is 2300 MeV
which is still higher than the N$\Xi$ threshold and so can not
form a narrow dibaryon resonance. The mass reduction due to the 
tensor interaction is very small as can be seen from Table III. 
Fig. 2 shows the S-D effective transition interactions in the
$\Sigma\Sigma$ channel due to $\pi$, K and gluon exchange
respectively. The $\pi$ contribution is reduced by about a factor 
of four. The gluon contribution is also very much reduced. In 
addition, it is repulsive and cancels the $\pi$ contribution. 
The K contribution is also repulsive and negligible. The $\eta$ 
contribution is even smaller.

The $\Lambda\Xi, \Xi\Xi$, and $\Xi\Omega$ states are near threshold. 
The tensor interactions in these channels are much less effective
than that in the deuteron channel and not strong enough to bind these
baryons into dibaryon resonances. The mass reductions due to the tensor
interactions in these channels are negligible as shown in Table III. 
Fig. 3 shows the S-D effective transition interactions in the $\Xi\Xi$ 
channel due to $\pi$, K and gluon respectively. The $\pi$ contribution 
is reduced even further than in the previous case. The gluon contribution 
is very much reduced in comparison with that in deuteron channel but a 
little bit enhanced in comparison with that in $\Sigma\Sigma$ channel. 
The K contribution is further verified to be negligible and $\eta$ 
negligible also.

Fig. 4 gives a direct comparison of the S-D effective transition 
interactions due to the tensor force of $\pi$ exchange in NN, 
$\Sigma\Sigma$ and $\Xi\Xi$ channels. Fig. 5 gives a direct 
comparison of the S-D effective transition interactions due to 
the tensor force of gluon exchange in NN, $\Sigma\Sigma$, $\Lambda\Xi$, 
$\Xi\Xi$ and $\Xi\Omega$ channels. The $\eta$ contribution has been 
calculated for all of these channels and all are negligible so they 
have not been shown explicitly.

As was mentioned in the introduction, the strangeness -3
$IJ^p=\frac{1}{2}2^+$ N$\Omega$ has also been predicted to be
tightly bound\cite{goldman,commun41,pang} if the tensor
interaction is neglected. Taking into account the tensor force,
this state can couple to D-wave $\Lambda\Xi$ and $\Sigma\Xi$
channels. Since $\Lambda\Xi$ is the lowest channel in all channels
with strangeness $S=-3$, here we only take into account this
channel. For the possible $N\Omega (IJ=1/2,2)$ bound state we
consider the two lowest $\Lambda\Xi$ D-wave decay channels: IS=1/2,0
and IS=1/2,1, respectively; here S specifies the channel spin. Such a tensor 
coupling has two effects: One is to modify the mass of the N$\Omega$ 
state; the other is to induce a transition from the N$\Omega$ to the 
D-wave $\Lambda\Xi$ final state and change the bound N$\Omega$ to a 
resonance with finite width. Both of these effects have been calculated.

The results show that, the mass of $N\Omega$(IJ=1/2,2) in the single
channel approximation is about 2566 MeV. Taking into account other 
S-wave channels coupling, such as $\Xi\Sigma, \Xi^*\Sigma,
\Xi\Sigma^*,\Xi\Lambda and \Xi^*\Sigma^*$, reduces the mass of the 
system to 2549 MeV, while adding the $\Lambda\Xi$ D-wave channel
coupling only changes the value of mass slightly (not more than 1 MeV).

The widths of $N\Omega$ decays to the $\Lambda\Xi$ D-wave with
different spins are listed in Table IV. For comparison, the width
of the $d^*$ decay to NN D-wave is also listed.

\vspace{12pt} Table IV. Decay width. I,S,J are isospin, spin and
total angular momentum, respectively.

\begin{tabular}{cc} \hline\hline
$N\Omega (IJ=1/2,2)\rightarrow \Lambda \Xi ~~D-wave
(S=0,I=1/2,J=2)$ &
~~~~~~~~$\Gamma = 12.0 keV $\\
$N\Omega (IJ=1/2,2)\rightarrow \Lambda \Xi ~~D-wave
(S=1,I=1/2,J=2)$ & ~~~~~~~~$\Gamma = 21.9 keV$ \\ \hline $d^*
(IJ=0,3) \rightarrow NN
~~D-wave (S=1,I=0,J=3) $ & ~~~~~~~~$\Gamma = 6.57 MeV $\\
\hline\hline
\end{tabular}

\vspace{12pt}
From Table IV we see that the width of $d^*$ decay to the NN
D-wave is 6.57 MeV. Comparison with our previous results\cite{prc62},
confirms that the width is not sensitive to the value of cutoff $r_0$.
The width of $N\Omega$ decays to $\Lambda\Xi$ D-wave is about tens of 
keV, about three orders of magnitude smaller. The result is not changed 
significantly for decay channels with different spin. For example,
the N$\Omega \rightarrow \Lambda\Xi$ (spin=0) D-wave decay
width is calculated to be 12.0 keV, and the N$\Omega \rightarrow 
\Lambda\Xi$ (spin=1) D-wave decay width is 21.9 keV. These results 
confirm our expectation that the N$\Omega$ is a narrow dibaryon 
resonance. (The width of N$\Omega \rightarrow \Lambda\Xi$ is smaller 
than that of the $d^* \rightarrow$ NN decay mainly due to the reduction 
of the tensor interaction, but also due to the fact that the N and 
$\Delta$ have the same flavor content while N, $\Omega$ and $\Lambda$, 
$\Xi$ have differing flavor content in each baryon.)

\section{Discussion and Conclusion}
The effects of the tensor interactions of gluon and Goldstone boson
exchanges on the dibaryon mass and decay width have been studied in 
the extended QDCSM. Only in the deuteron channel is the tensor 
interaction of $\pi$ exchange strong enough to bind the two nucleons 
into a loosely bound state. No other near threshold six-quark 
state studied so far in the u, d, s three flavor world can be bound 
together by the additional attraction induced by these tensor 
interactions. The S-D wave effective transition interactions due to 
$\pi$ and gluon tensor forces both decrease rapidly with increasing 
strangeness. In the $\Sigma\Sigma$ and $\Xi\Xi$ channels the effective 
transition interactions due to the gluon tensor term become repulsive 
and cancel the attractive $\pi$ contribution. The tensor contribution
of explicit K and $\eta$ meson is confirmed as negligible due to the 
same short range truncation as for the $\pi$, in our model 
approach\cite{pang,commun38}

The mass shift of the $IJ^p=\frac{1}{2}2^+$ N$\Omega$ state
induced by the tensor interaction is small (not more than 1 MeV)
and the D-wave partial decay widths to $\Lambda\Xi$ with spin-0 and
spin-1 are only 12.0 and 21.9 keV, respectively. Hence, the 
$IJ^p=\frac{1}{2}2^+$ N$\Omega$ state appears to be a good 
candidate for a narrow dibaryon resonance. Altogether there are 
only two promising narrow dibaryon resonances in the light flavor 
world in our model approach: The $IJ^p=03^+$ $d^*$ and the 
$IJ^p=\frac{1}{2}2^+$ N$\Omega$.

The $H$ particle and di-$\Omega$ are marginally strong interaction
stable in our model. However the theoretical binding energies of
both are small (only few MeV\cite{commun38,pang}). Table II shows
that the calculated ground octet and decuplet baryon masses
deviate from the measured ones about 18 MeV on average. A reasonable
estimate of the model uncertainty for the dibaryon mass would be
at least that large. Therefore, in our model, it is unjustified to 
assert that the $H$ particle and di-$\Omega$ might be strong interaction 
stable dibaryon candidates. This estimation is consistent with the
latest di-$\Lambda$ hypernuclear findings\cite{tak}. There are
various broad resonances with widths $\sim 150 \rightarrow 250$
MeV around the $d^*$ mass ($\sim$ 2180 MeV) in our model which
makes the analysis of the NN scattering more difficult in the energy
region 2.1 $\rightarrow$ 2.4 MeV where a broad bump has been found
in the pp and np total cross sections. We will report on those
results later. The $SIJ=-3,1/2,2$ N$\Omega$ state is quite
convincingly lower in mass than the N$\Omega$ threshold, and
quite possibly lower than the $\Lambda\Xi\pi$ threshold, as well.
We have shown the decay width to be as small as tens of keV. Such
a narrow dibaryon resonance might be detected by reconstructing
the invariant mass of its two body decay products, $\Lambda$ and 
$\Xi$, in high $\Omega$ production reactions using RHIC at Brookhaven 
and COMPASS at CERN and the future ones at JHF in Japan and FAIR 
in Germany.

This model, the extended QDCSM, which proposes a new mechanism to
describe the NN intermediate range attraction instead of the
$\sigma$ meson, well describes, with the fewest parameters, the
properties of the deuteron and the existing NN, N$\Lambda$ and
N$\Sigma$ scattering data. Up to now, it is the only model
which gives an explanation of the long standing fact that the
nuclear and molecular forces are similar in character despite the
obvious length and energy scale differences and that nuclei are
well described as collection of A nucleons rather than 3A quarks.
In view of the fact that the $H$ particle has not been observed
experimentally, the BB interaction in the $\Lambda\Lambda$
channel\cite{tak} predicted by this model may be a good
approximation of the real world.  Based on these facts we suppose
the predictions about dibaryon states of this model might also be
approximately correct. Of course, the QDCSM is only a model of QCD. 
The high spin, high strangeness dibaryon resonance,
$IJ^p=\frac{1}{2}2^+$ N$\Omega$, may be a good venue to search for
new hadronic matter and to test whether or not the QDCSM mechanism
for the intermediate range attraction is realistic.

This work is supported by NSFC contracts 90103018 and 10375030.
and by the U.S. Department of Energy under contract W-7405-ENG-36.
F. Wang would like to thank the ITP for their support through the 
visiting program.

\pagebreak
FIGURE
CAPTIONS

Fig.1 The effective S-D wave transition interactions of gluon and $\pi$
tensor force in the deuteron channel.

Fig.2 The effective S-D wave transition interactions of gluon and $\pi$,
K tensor force in the $IJ^p=11^+$  $\Sigma\Sigma$ channel.

Fig.3 The effective S-D wave transition interactions of gluon and $\pi$,
K tensor force in the $IJ^p=01^+$  $\Xi\Xi$ channel.

Fig.4 A comparison of effective S-D wave transition interactions of
$\pi$ tensor force in NN, $\Sigma\Sigma$, $\Xi\Xi$ channels.

Fig.5 A comparison of effective S-D wave transition interactions of
gluon tensor force in NN, $\Sigma\Sigma$, $\Lambda\Xi$, $\Xi\Xi$,
$\Xi\Omega$ channels.

\pagebreak

\begin{figure*}[t]
\includegraphics[height=4.0in]{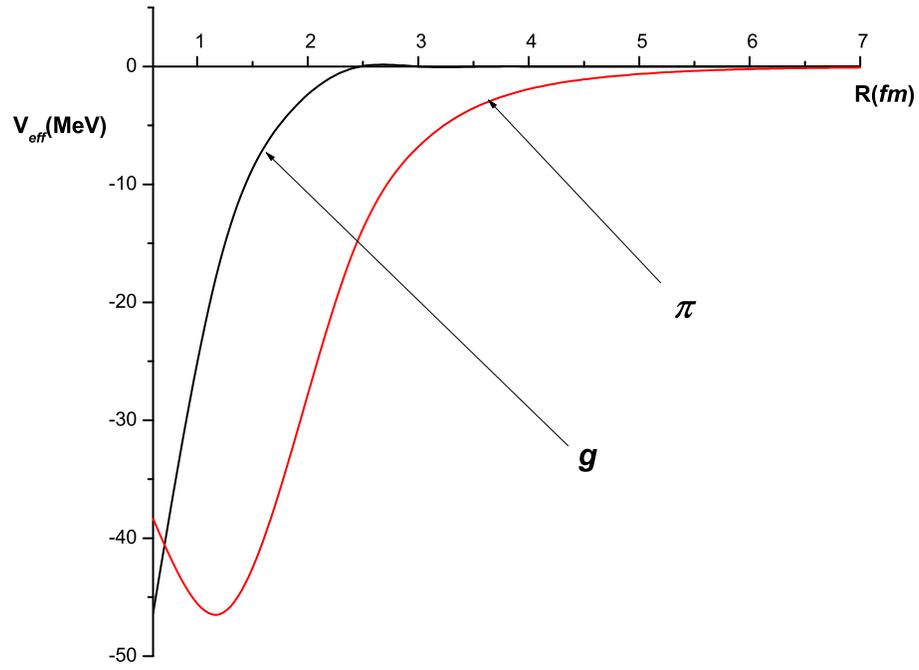}
\caption{The effective S-D wave transition interactions of gluon and $\pi$
tensor force in the deuteron channel.}
\label{FIG1}
\end{figure*}

\begin{figure*}[t]
\includegraphics[height=4.0in]{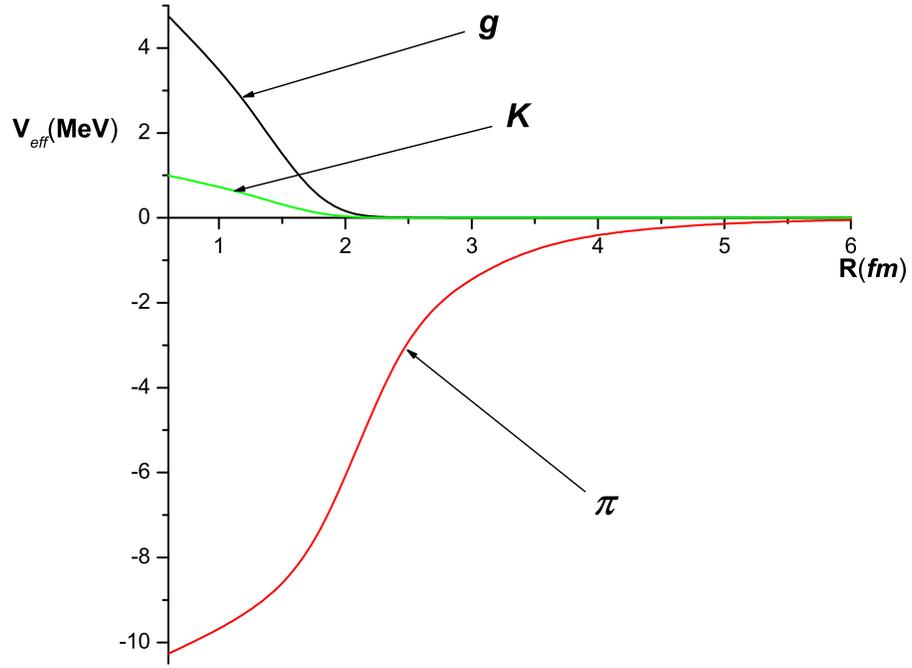}
\caption{The effective S-D wave transition interactions of gluon and $\pi$,
K tensor force in the $IJ^p=11^+$ $\Sigma\Sigma$ channel.}
\label{FIG2}
\end{figure*}

\begin{figure*}[t]
\includegraphics[height=4.0in]{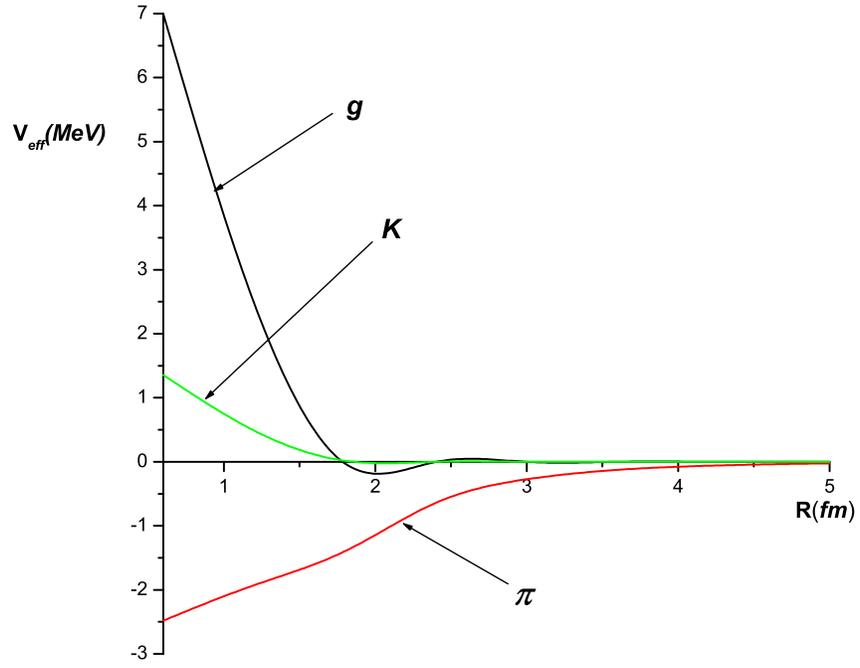}
\caption{The effective S-D wave transition interactions of gluon and $\pi$,
K tensor force in the $IJ^p=01^+$ $\Xi\Xi$ channel.}
\label{FIG3}
\end{figure*}

\begin{figure*}[t]
\includegraphics[height=4.0in]{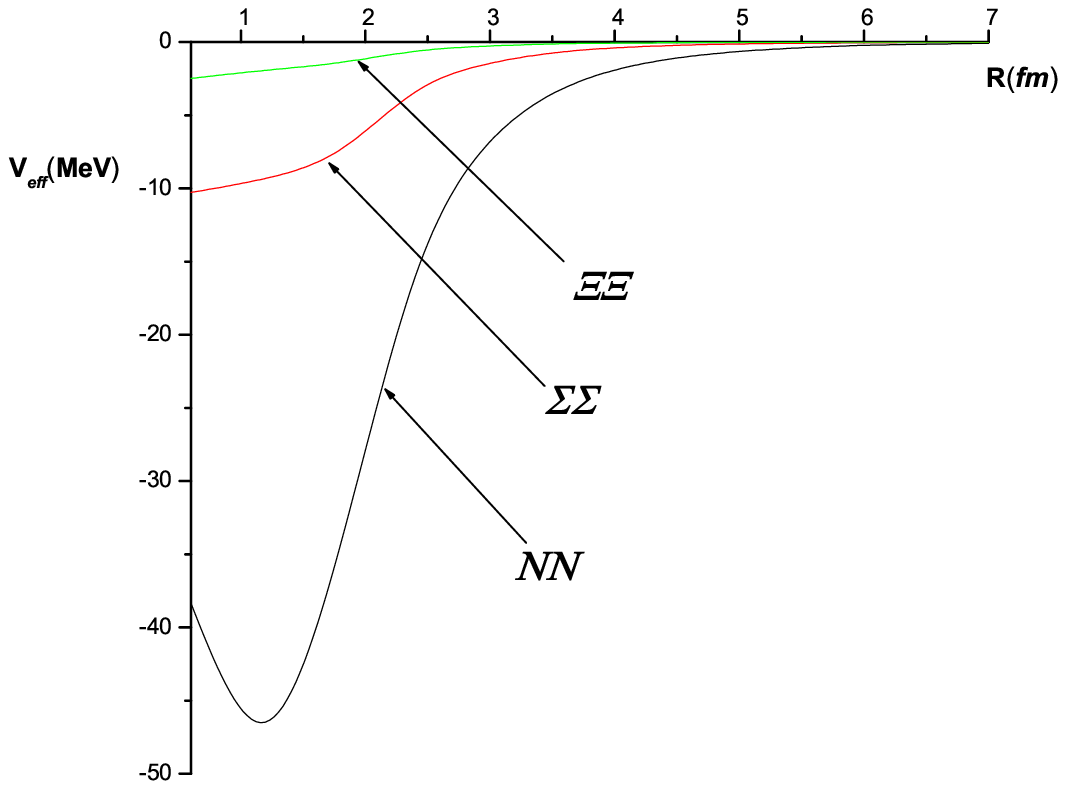}
\caption{A comparison of effective S-D wave transition interactions of
$\pi$ tensor force in NN, $\Sigma\Sigma$, $\Xi\Xi$ channels.}
\label{FIG4}
\end{figure*}

\begin{figure*}[t]
\includegraphics[height=4.0in]{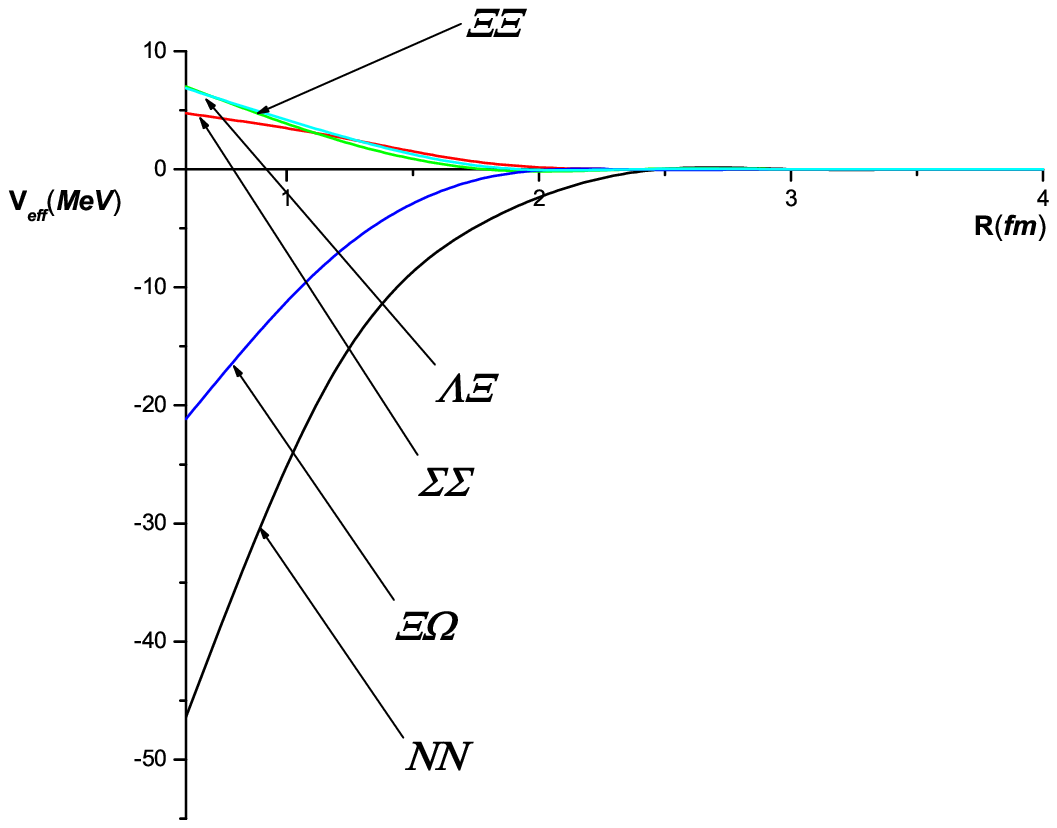}
\caption{A comparison of effective S-D wave transition interactions of
gluon tensor force in NN, $\Sigma\Sigma$, $\Lambda\Xi$, $\Xi\Xi$,
$\Xi\Omega$ channels.}
\label{FIG5}
\end{figure*}

\end{document}